\documentclass[preprint]{aastex}
\usepackage{epsfig,emulateapj5}

\newcommand{\htwo}{H$_2$}
\newcommand{\rv}{$R_V$}
\newcommand{\ebv}{$E(B-V)$}

\shortauthors{BURGH ET AL.}
\shorttitle{FAR-UV NON-LINEAR EXTINCTION AND CO ABSORPTION}

\begin{document}
\makeatletter
\newenvironment{tablehere}
  {\def\@captype{table}}
  {}

\newenvironment{figurehere}
  {\def\@captype{figure}}
  {}
\makeatother

\slugcomment{To appear in the Astrophysical Journal, September 20, 2000}

\title{On the Correlation Between CO Absorption and Far-Ultraviolet
Non-Linear Extinction Toward Galactic OB Stars}

\author{Eric B. Burgh, Stephan R. McCandliss, B-G Andersson, and Paul
D. Feldman}

\affil{Department of Physics and Astronomy, The Johns Hopkins
       	University, Baltimore, MD 21218}
\email{ebb@pha.jhu.edu,stephan@pha.jhu.edu,bg@pha.jhu.edu,pdf@pha.jhu.edu}

\received{}
\revised{}
\accepted{}

\ccc{}
\cpright{AAS}{}

\begin{abstract} 
A sample of 59 sight lines to reddened Galactic OB stars was examined
for correlations of the strength of the CO Fourth Positive ($A^1\Pi -
X^1\Sigma^+$) absorption band system with the ultraviolet interstellar
extinction curve parameters.  We used archival high-dispersion NEWSIPS
IUE spectra to measure the CO absorption for comparison to parametric
fits of the extinction curves from the literature.  A strong correlation
with the non-linear far-UV curvature term was found with greater
absorption, normalized to \ebv, being associated with more curvature.  A
weaker trend with the linear extinction term was also found. Mechanisms
for enhancing CO in dust environments exhibiting high non-linear
curvature are discussed.
\end{abstract}


\section{Introduction}

The extinction of starlight in the ultraviolet (UV; $\lambda\lesssim3200$
\AA) can be characterized by three main components: a linear rise, a
Lorentzian-like bump centered on $\lambda\sim2175$ \AA, and a
far-ultraviolet ($\lambda\lesssim1700$ \AA) non-linear rise.  The
general features of the extinction curve can be understood in terms of
dust grain populations \citep[e.g.]{Mathis77,Draine84}.
\citet{CCM88,CCM89} showed that the overall shape of the UV extinction
curve can be estimated by the single parameter \rv [$\equiv
A(V)/E(B-V)$].  However, individual sight lines exhibit large deviations
from the average extinction curve, particularly in the far-UV rise
\citep{Mathis92}.  A detailed understanding of these deviations can
provide information about the nature of the particles responsible for
the extinction, putting constraints on interstellar grain
models.

A number of studies have investigated the possible correlations between
the various components of the UV extinction curve and gas phase
abundances \citep{Joseph89,Jenniskens92}, infrared emissions
\citep{Cox87,Hackwell91,Boulanger94}, very broadband structure
\citep{Jenniskens94}, and diffuse interstellar band (DIB) absorptions
\citep{Desert95}.  The \citet{Jenniskens92} study was the first to
positively identify a correlation of the strength of the non-linear rise
with the abundance of a gas phase atom or molecule, finding that a
larger column of CH implies a stronger far-UV non-linear rise, and the
inverse relationship for CH$^+$.

No such study has to date been performed for CO, a molecule of great
astrophysical importance. Being the second most abundant molecule after
\htwo, it is used as a tracer of \htwo\ gas and an indicator of
the total mass of molecular clouds
\citep{Scoville87,Bloemen89,Hunter94}.  Having allowed rotational
transitions that are readily observed in emission at radio wavelengths,
the CO molecule is commonly used to probe dark molecular clouds.  CO
also has an extensive band system in the ultraviolet, which can be used
to investigate the molecular content of diffuse and translucent clouds
when observed in absorption toward background stars.

The wavelength coverage of the short-wavelength prime (SWP) camera of
the \textit{International Ultraviolet Explorer} (IUE) satellite, 1150 --
2000 \AA, spans as many as 19 bands of the Fourth Positive ($A^1\Pi -
X^1\Sigma^+$) ($v'$--0) band system beginning with the (0--0) band at
1544.5 \AA\ and continuing to shorter wavelengths.  For this study, we
searched the IUE archival data for ultraviolet absorption of CO along 59
lines of sight toward Galactic O and B stars for which UV extinction
curves have already been published, and parameterized as described
below, to investigate the possible correlations with the various curve
features.

\citet[FM hereafter]{FM90} have proposed a parameterization that
accurately fits the extinction curves with a single analytical
expression using six parameters.  Using this method, the extinction
curve, normalized to unit \ebv\ and with $x\equiv\lambda^{-1}$, can be
expressed as
\begin{equation}
k(x)\equiv\frac{E(\lambda-V)}{E(B-V)}=c_1+c_2x+c_3D(x;\gamma,x_0)+c_4F(x)
\label{eq1}
\end{equation}
where
\begin{equation}
D(x;\gamma,x_0)=\frac{x^2}{(x^2-x_0^2)^2+x^2\gamma^2}
\end{equation}
is the Lorentzian-like ``Drude'' profile, representing the 2175 \AA\
bump with $x_0$ the bump peak and $\gamma$ its width, and
\begin{equation}
F(x)= \left\{ \begin{array}{ll}0.53(x-5.9)^2+0.05446(x-5.9)^3 & x > 5.9
 \mu m^{-1} \\ 0 & x \leq 5.9 \mu m^{-1} \end{array} \right.
\end{equation}
is the far-UV curvature term, whose strength is determined by a single
parameter, $c_4$.

\section{The Sample}

Our sample consists of 59 reddened stars ($0.24\leq E(B-V)\leq1.09$).
The UV extinction curves for these stars have been previously determined
using the pair method by FM and \citet{Aiello88}.  The extinction curve
parameters for stars from the Aiello et al. sample are taken from
\citet[JG hereafter]{Jenniskens93}, who used the FM fitting routine.
For stars that appear in both samples, we adopt the FM numbers.  The
parameters from JG show a small (1-$\sigma$) systematic offset from the
FM data, producing slightly smaller $c_2$ values and larger $c_4$
values, approximately $-0.1$ and $+0.1$ respectively. These systematic
errors do not significantly affect the results of the analysis.

Of the stars in the FM and JG samples, only those for which
high-dispersion short-wavelength IUE spectra exist are considered here.
We also consider only stars of spectral type B5 and earlier and
luminosity classes III and V, since spectral mismatch using the pair
method can strongly affect the extinction curve fitting at later types
and higher luminosities \citep{Massa83}.  Although these stars do not
represent a purely random sampling of interstellar sight lines, they do
sample different interstellar environments, including dense and diffuse
media, as well as regions of recent early-type star formation.

\section{Analysis}

For each star high-dispersion short-wavelength New Spectral Image
Processing System (NEWSIPS) IUE spectra were obtained from the archive
maintained by the Astrophysics Data Facility (ADF).  The quality of the
data varies strongly from object to object due to the differences in
exposure time, brightness of the source, and number of spectra
available.  To improve the signal-to-noise per source, multiple spectra
were co-added by weighted mean when available.  However, no further
processing was performed on the data.  Figure \ref{fig1} shows
examples of both strong and weak CO absorption.

\placefigure{fig1}

We measured the equivalent width of the (2--0) band in each spectrum.  A
linear continuum was fit and the equivalent width measured by
integrating over the absorption line profile (typically a span of
$0.4-1$ \AA).  The 1-$\sigma$ error was computed by summing the
normalized errors for each pixel in quadrature.  For non-detections, the
upper limit was taken to be the error computed across 0.4 \AA.  The
measured equivalent widths and the IUE exposure numbers for each star
are listed in Table \ref{table1}.

\placetable{table1}

The (2--0) band, located near 1477.6 \AA, was chosen for several reasons.
It is well separated from the nearby $^{13}$CO band at 1478.8 \AA\, and
is relatively uncontaminated by other interstellar or stellar features
for all the spectral types considered in this study.  It is also the
least susceptible to background subtraction errors known to occur in
NEWSIPS data because it lies near the sensitivity peak of the echelle
spectrograph \citep{Massa98}.  It also has the highest band oscillator
strength of the Fourth Positive band system and thus is the most likely
to be observed in low $S/N$ data.  

The resolution of the IUE echelle spectrograph ($R\approx10^4$) combined
with the limiting $S/N$ ($\approx 20:1$) of the SEC vidicon detectors
\citep{Cowie86} conspires to restrict our typical 2-$\sigma$ equivalent
width detection limit to $\simeq$ 30~m\AA.  As we will show in the
Appendix, even at this equivalent width, all absorption lines are
affected by saturation and thus the conversion of the molecular band
equivalent width to a corresponding column density depends on
assumptions of the gas rotational excitation temperature, $T$, and doppler
velocity, $b$.  Although the various sight lines are likely to have different
values for $T$ and $b$, on average the values for these parameters lie
around $T \approx 4$ K and $b \approx1$~km~s$^{-1}$ \citep{Gredel91}, as
observed in higher quality data.  We will adopt these values to convert
the equivalent width to column density for the comparison to previous
work and allow a statistical assessment of the robustness of our result
with respect to the phenomenologically more meaningful column density.

\begin{figure}
\centerline{\epsfig{figure=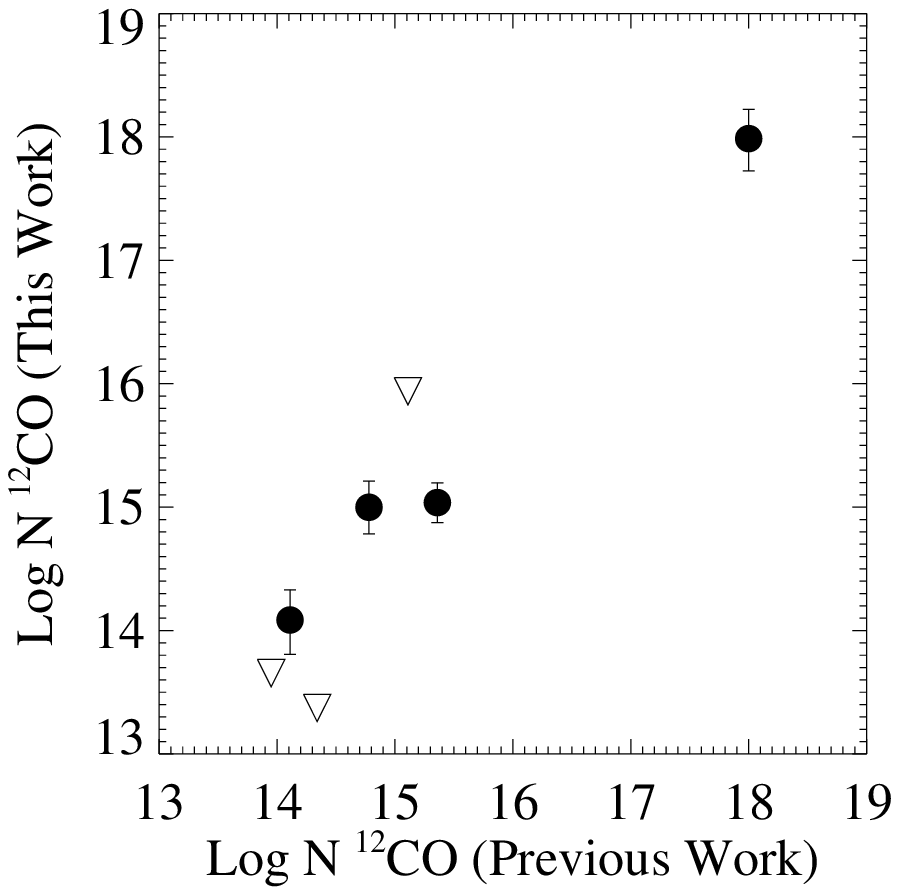}}
\caption{Log-log plot of $^{12}$CO column density in
cm$^{-2}$ comparing our data to that listed in Federman et~al. (1994)
for those sight lines in common.  Column densities were determined using
a curve-of-growth method assuming a doppler parameter $b=1$ km s$^{-1}$
and rotational excitation temperature $T=4$ K.  Upper limits
(2-$\sigma$) are shown as open triangles.
\label{fig2}}
\end{figure}

For sight lines whose CO absorption has been previously studied our
measurements agree very well with the published data. In particular, the
equivalent widths for the (2--0) bands of HD~21483, HD~47129, and
HD~149757 of $322\pm27$ \citep{Joseph86}, $46\pm17$ \citep{Tarafdar82},
and $74\pm7$ \citep{Wannier82} all agree within the error with our
measured values.  Four more sight lines (HD~34078, HD~37903, HD~147933,
and HD~193322) have been observed in other bands, either higher $A-X$
bands or the shorter wavelength $B-X$ and $C-X$ bands.  Column densities
for all of these stars are compiled in \cite{Federman94}.  By converting
our measured equivalent width to column density as described in the
Appendix we directly compare our results for all the aforementioned
sight lines in Figure~\ref{fig2}.

\placefigure{fig2}

For the purpose of comparison to the extinction curve parameters as
specified in Equation~\ref{eq1}, the
equivalent width is normalized to unit reddening.  This produces a
measure of the relative abundance of CO to the total amount of matter
along the line of sight. Figure~\ref{fig3} shows the correlation plots
of this quantity versus each of the six extinction curve parameters for
the stars listed in Table~\ref{table1}.

\placefigure{fig3}

These data were analyzed using the Astronomy SURVival Analysis (ASURV)
software package Rev 1.2 \citep{Isobe90,LaValley92}, which implements
the survival analysis methods presented in \citet{Feigelson85} and
incorporates the upper limits in a statistically appropriate manner.
For each parameter of the extinction curve, the Spearman's Rho value, a
non-parametric rank-order correlation coefficient, was computed.  This
value, which can range between $1$ and $-1$ for a perfect correlation and
anti-correlation respectively, was tested for its significance by
calculating a $t$-value, which is distributed approximately as Student's
distribution with $N-2$ degrees of freedom and is effectively a measure of the
standard deviations from the null hypothesis.  This allows the probability
that the correlation could be drawn randomly from this distribution to
be determined.  These results are tabulated in Table \ref{table2}.

\placetable{table2}

\section{Results and Discussion}

\begin{figure}
\centerline{\epsfig{figure=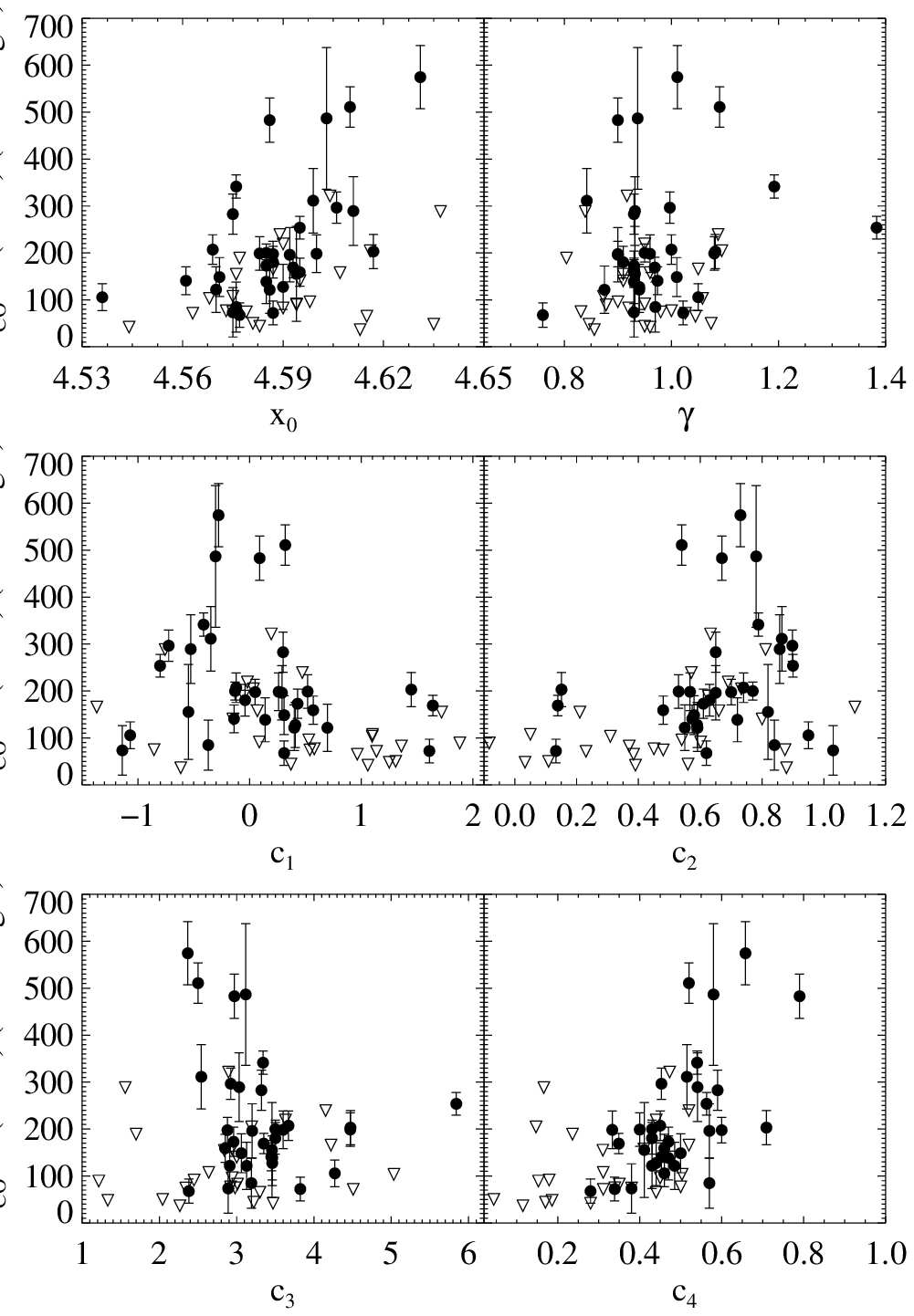}}
\caption{Correlation diagrams for the normalized equivalent
width of the CO $A-X$ (2--0) band in m\AA\ per magnitude of \ebv.  Upper
limits (2-$\sigma$) are shown as open triangles.
\label{fig3}}
\end{figure}

No significant correlations were seen with the extinction curve
parameters $x_0$, $\gamma$, or $c_3$ (see Figure \ref{fig3}).  The
relationship with $c_2$ (middle right panel of Figure \ref{fig3}), while
not showing a linear correlation does show an interesting
bifurcation. (FM note that $c_1$ and $c_2$ are strongly correlated and
thus only the relationship with $c_2$ is considered here.)  Although a
high $c_2$ does not guarantee a large CO abundance, the largest
abundances are seen when $c_2 > 0.5$.  Of notable exception is
HD~147889. This highly reddened star ($E(B-V)=1.09$) has a $c_2$ of
0.151 but the absorption suggests a column density as high as
$3\times10^{17}$ cm$^{-2}$.  However, it must be noted that this star
has a high $c_4$ value of 0.709.  As $c_2$ is inversely proportional to
\rv\ \citep{Fitzpatrick99}, our result would suggest that lower CO
abundances would be observed along sight lines with \rv $\gtrsim 3.6$.
\citet{Cardelli88} noted a decline in the abundances of \htwo\ and CH
with increasing \rv\ and a similar break in the CH abundance at \rv =
3.5 for clouds with $A_V<2$.  The interstellar average \rv\ is 3.1,
which corresponds to a $c_2$ of 0.7, and thus it is apparent that the
slope of the linear rise is not the dominant factor determining the
abundance of CO in average clouds.

\begin{figure}
\centerline{\epsfig{figure=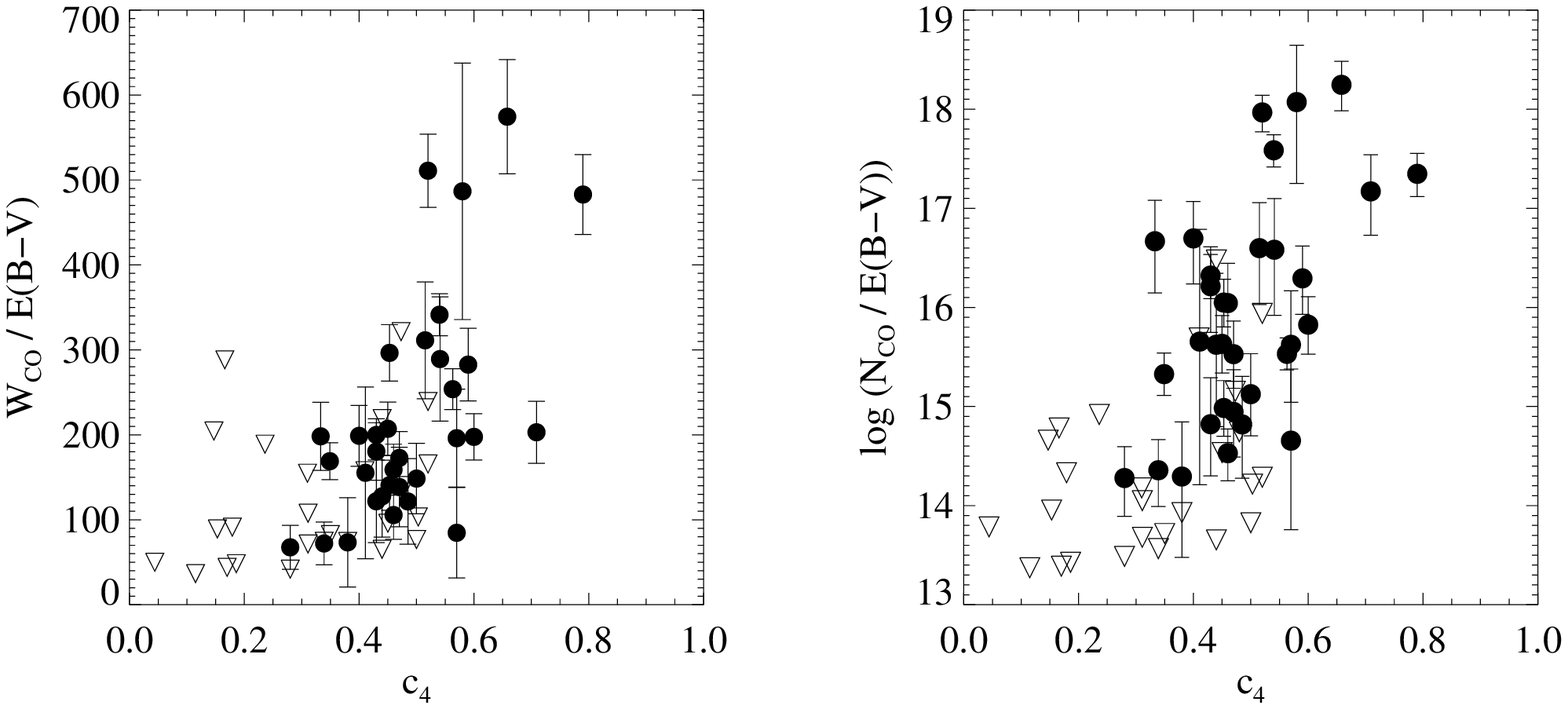}}
\caption{Left panel: Correlation diagram for the normalized
equivalent width in m\AA\ per magnitude of \ebv\ versus $c_4$.  Upper
limits (2-$\sigma$) are shown as open triangles.  Right panel:
Correlation diagram for column density in cm$^{-2}$ per magnitude assuming $T$
= 4 K and $b$ = 1 km~s$^{-1}$.
\label{fig4}}
\end{figure}

The plot with $c_4$, the strength of the non-linear far-UV rise, (left
panel of Figure \ref{fig4} -- same as bottom right panel of Figure
\ref{fig3}) shows the CO abundance being correlated with stronger far-UV
curvature.  We note that for $c_4 \gtrsim 0.5$ we always detect CO
and only upper limits exist for $c_4$ below 0.25.  This suggests that
the carrier of the non-linear rise has a more direct association with
the CO abundance than that of the linear rise ($c_2$).  The right panel
of Figure \ref{fig4} shows the relationship between the CO column
density per \ebv\ and $c_4$ assuming $b$ = 1 km~s$^{-1}$ and $T$ = 4 K.
Although these values are likely not correct for each individual star it
demonstrates the effective range of column densities observed and the
lower limit of detectability by IUE.  This correlation produces a
Spearman's Rho of 0.630, consistent with the equivalent width
relationship.

\placefigure{fig4}

The abundance of CO along a given sight line is governed by the balance
between formation through chemical networks involving species such as
C$^+$, OH, and CH, and destruction through photodissociation
\citep{vDB88,Federman89}.  The correlation of CO with $c_4$ suggests
that the presence of the carrier of the non-linear far-UV extinction is
conducive to the presence of CO.  This can be accomplished either by not
destroying CO through shielding of the photodissociating flux or by
aiding in the formation of CO by not photodissociating the precursor
molecules.

\begin{figure}
\centerline{\epsfig{figure=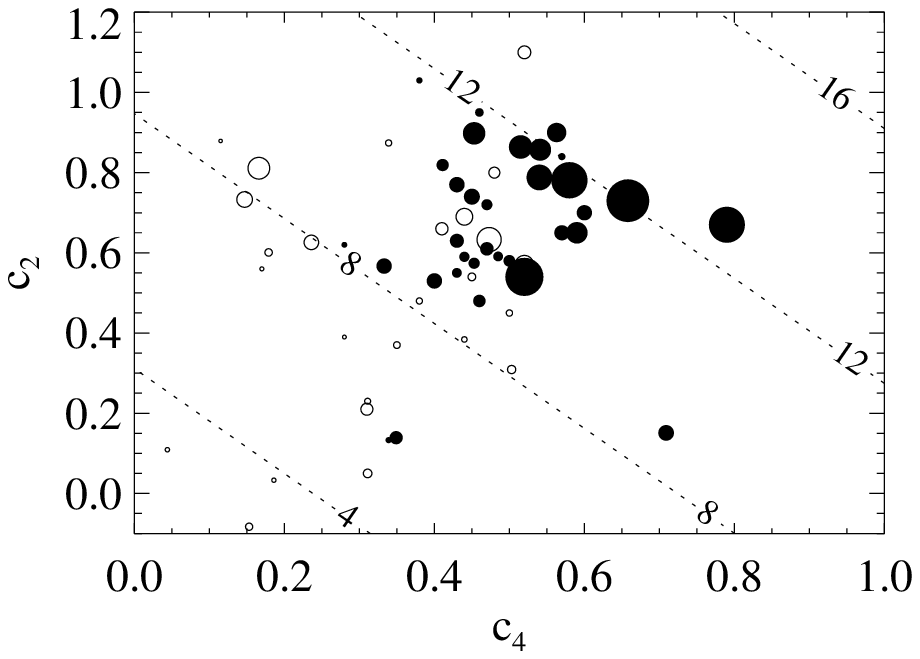}}
\caption{Plot of $c_2$ v. $c_4$. The size of the circle
indicates the strength of the CO absorption.  Open circles are 
2-$\sigma$ upper limits.The dotted lines correspond to constant $k(x)$
for $\lambda=1076$ \AA, with increasing extinction to the upper right.
\label{fig5}}
\end{figure}

CO photodissociates through the discrete absorptions of far-UV photons
resulting in the excitation of the molecule into predissociating states.
At the edges of clouds, this occurs primarily through the $E-X$ (0--0)
band at 1076~\AA.  However, as optical depth increases into the cloud,
most of the photodissociation moves to shorter wavelength transitions
\citep{vDB88}, where the extinction of flux by the non-linear rise
increases dramatically.  \citet{vDB89} have demonstrated that the
photodissociation rate of CO is reduced in cloud models that use steeper
extinction curves to model the continuum extinction of far-UV flux. It
is therefore possible that we are observing an increase in molecular
content associated with a decrease in photodestruction since the
strength of the parameters $c_2$ and $c_4$ regulate the far-UV flux at
short wavelengths. This is suggested in Figure \ref{fig5}, where we show
the relation of these parameters to the CO normalized equivalent width.
However, it is important to realize that the actual attenuation of
radiation in a cloud depends strongly on the far-UV scattering
characteristics of the grains responsible for the extinction, even when
the extinction rises sharply in the far-UV \citep{Flannery80}.

\placefigure{fig5}

Although Figure \ref{fig5} is suggestive, it is far from conclusive.  If
the carrier of the non-linear rise aids in the formation of CO, or any
of its precursor molecules, such as CH and OH, the CO formation rate may
be enhanced enough to explain the rapid increase in column densities
seen in Figure \ref{fig4}. 

\vspace{.2in}
\section{Conclusions}

The abundance of CO, as measured by the equivalent width of the UV
absorption of the $A-X$ (2--0) band normalized to unit \ebv, is
correlated with the strength of the non-linear far-UV rise in the UV
extinction curves toward Galactic O and B stars.  This correlation is
indicative of either a decrease in the photodestruction rate of CO with
increased extinction or of a dust environment that is conducive to the
enhanced formation of CO or its precursor molecules.  The linear rise
does not correlate with CO abundance, although for sight lines with
large \rv\ ($\gtrsim 3.6$) we observe mostly small columns
($\lesssim10^{15}$ cm$^{-2}$ per \ebv).
These results, together with those of \citet{Cardelli88} reinforces the
notion that the use of the visual extinction, $A_V$, as the independent
variable in global studies of interstellar chemistry has to be viewed
with some caution.  

Future study of these relationships could benefit from a quantitative
determination of the CO column density and an exploration of the
\htwo/CO ratio in these environments.  For \htwo, the ideal instrument is
the recently launched \textit{Far-Ultraviolet Spectroscopic Explorer}
(FUSE) satellite.  FUSE will be able to measure the \htwo\ column from
the individual lines of the Werner and Lyman band systems, as well as
observe the absorptions from the CO Hopfield-Birge bands.  However, more
accurate CO observations could be made by the Space Telescope Imaging
Spectrograph (STIS) for two reasons.  The wavelength separation of the
rotational lines within the $A-X$ bands is greater than that of the
$B-X$ and $C-X$ and the STIS Echelle modes have higher resolution than
FUSE, allowing for the measurement of the ground state rotational
temperature and doppler velocity, and therefore a precise derivation of
column density.  Finally, we note that the ratio of true absorption to
scattering by dust in the far-UV is critical to the interpretation of
these results.  Studies aimed at investigating this relation in
conjunction with the FM extinction parameterization would be useful.
Such measurements would facilitate a more detailed analysis of the
conditions in these cloud environments.

\acknowledgments All IUE data were obtained from the Astrophysics Data
Facility at NASA's Goddard Space Flight Center.  The authors would like
to thank Derck Massa for valuable discussions.  We also thank the
referee, Steve Federman, for several helpful comments and
suggestions.  This work was supported by NASA grant NAG5-5122 to The
Johns Hopkins University.

\appendix

Typically, the column density is derived from the measured equivalent
width through the use of an appropriate curve-of-growth.  For molecules,
the shape of a band integrated curve-of-growth depends strongly on the
rotational excitation temperature, $T$, of the molecules and the
line-broadening doppler parameter, $b$.  Unfortunately, the spectral
resolution of the IUE data is not high enough to allow for a direct
measurement of the $b$ parameter and only in the cases of the strongest
absorptions is it possible to put limits on the rotational temperature
from the observed absorption profile.

\begin{figure}
\centerline{\epsfig{figure=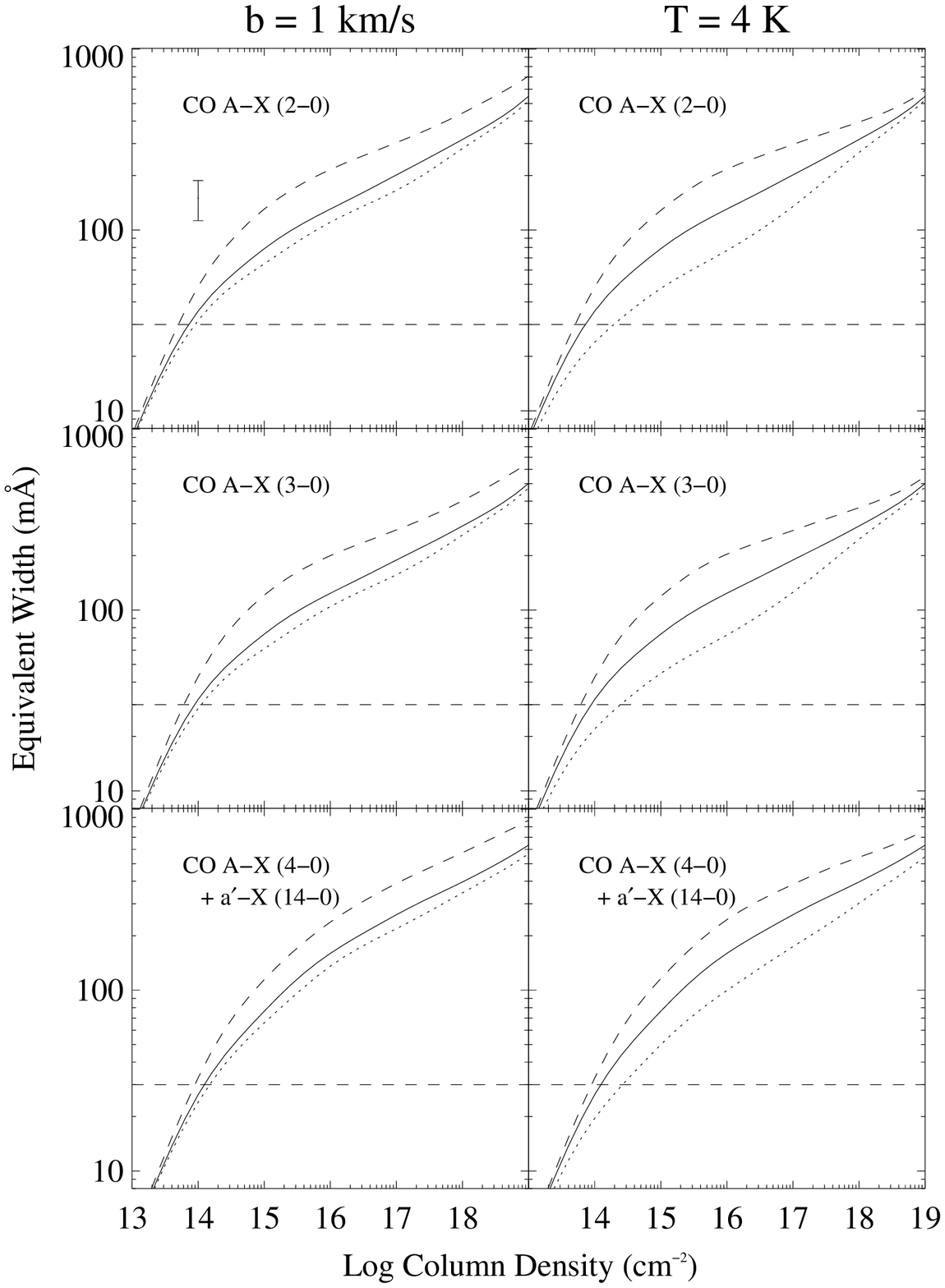}}
\caption{Curves-of-growth for the CO $A-X$ (2--0),
(3--0), and (4--0) bands.  The (4--0) band is blended with the $a'-X$ (14--0)
band.  The left panels are for $b=1$ km s$^{-1}$ with dashed, solid and
dotted lines corresponding to a rotational temperature of $T$ = 10, 4, and
3 K respectively.  The right panels are for $T$ = 4 K with dashed, solid
and dotted lines corresponding to a doppler parameter $b$ = 2, 1, and
0.5 km s$^{-1}$ respectively.  A horizontal dashed line shows our
typical 30~m\AA\ detection limit.  The top left panel shows a typical
error bar in equivalent width.
\label{fig6}}
\end{figure}

The use of multiple bands could produce a more accurate column density
determination by limiting the possible values of $b$ and $T$.  However,
to fit the relative amounts of absorption in multiple bands with a
single model based on three parameters ($N$, $b$, and $T$) it is
desirable to have accurate equivalent widths of more than three bands.
The (0--0) and (1--0) bands of the $A-X$ band system have blended
$^{12}$CO and $^{13}$CO components.  The (3--0) and (4--0) bands can
sometimes be measured although the continuum has definite structure for
some of the spectral types of stars considered in this study, while the
(5-0) band is blended with the \ion{Si}{4} $\lambda1394$ feature.  Thus,
the measurement of the equivalent width of these bands is dependent on
an accurate fit to the continuum in these regions.  The (6-0) band is
not usually blended with other features, but it has an oscillator
strength five times weaker than the (2--0) band and is only observed for
the largest column density sight lines.  Figure \ref{fig6} shows band
integrated curves-of-growth for the (2--0), (3--0), and (4--0) bands of
the CO $A - X$ band system computed using the band oscillator strengths
and line wavelengths of \citet{MortNor}. The (4--0) band is blended with
the $a'-X$ (14--0) band due to quantum mechanical mixing of these
states.  All of these bands show the effects of saturation above the IUE
detection limit.  As can be appreciated from the following example,
attempts to fit the equivalent widths of multiple bands can produce
non-unique solutions to $N$, $b$, and $T$.  A typical measurement
(HD~192281 e.g.) yields $W_{(2-0)} = 140 \pm 13$~m\AA, $W_{(3-0)} = 130
\pm 16$~m\AA, and $W_{(4-0)} = 94 \pm 13$~m\AA.  Utilizing the
curves-of-growth shown in Figure \ref{fig6} we find that within the
range of reasonable $b$-values and rotational excitation temperatures
the uncertainty in column density is approximately two orders of
magnitude.

\placefigure{fig6}

Another approach is to synthesize a model spectrum that can be compared
directly to the data.  We attempted to solve for $N$, $b$, and $T$ by
generating a grid of models with various values for these parameters and
minimizing the $\chi^2$ value determined.  This method proved successful
only for those sight lines with the highest column density ($\sim
10^{17}-10^{18}$ cm$^{-2}$).  The column density derived this way was
usually within a factor of 2 or 3 of the valued determined by assuming
$T = 4$ K and $b = 1$ km s$^{-1}$.  However, for smaller column sight
lines ($\sim 10^{16}$ cm$^{-2}$) the column derived differed by as much
as an order of magnitude.  This is due to the fact that the absorption
profiles for these columns at the resolution of IUE strongly mix the
effects of differing $T$ and $b$, especially at low $T$.  These results
suggest that the assumption of average values for $T$ and $b$ may be
justified for a statistical comparison to the extinction curve
parameters, but may not produce reliable answers on an individual basis.

\newpage

\begin{deluxetable}{llcccccl}
\tablewidth{7in}
\tablecaption{Stellar extinction curve and CO absorption information}
\tablehead{&  \colhead{Sp.} & & & & 
\colhead{$W_\mathrm{CO}$ (2--0)\tablenotemark{a}} & &\\
\colhead{Star} & \colhead{Type} & \ebv & \colhead{$c_2$} &
\colhead{$c_4$} & \colhead{(m\AA)} & \colhead{Ref.} &
SWP number(s)\tablenotemark{b} }
\startdata
HD 13268 & O8V	& 0.44 & \phm{-}1.03\phn & 0.38\phn & \phn32 $\pm$ 23 & 1 & 09323 \\
HD 14434 & O6V	& 0.48 & \phm{-}0.80\phn & 0.48\phn & $<$ 67 & 1 & 16094 \\
HD 14442 & O6III& 0.73 & \phm{-}0.69\phn & 0.44\phn & $<$ 160 & 1 & 53141\\
HD 15558 & O5III& 0.83 & \phm{-}0.59\phn & 0.44\phn & 106 $\pm$ 40 & 1 & 08322s \\
HD 15629 & O5V	& 0.75 & \phm{-}0.63\phn & 0.43\phn & 135 $\pm$ 25 & 1 & 10754 \\[.05in]
HD 21483 & B3III& 0.55 & \phm{-}0.730 & 0.658 & 316 $\pm$ 37 & 2 & 38298\\
HD 34078 & 09.5V& 0.53 & \phm{-}0.571 & 0.520 & $<$ 126 & 2 & 22108, 37429\\
HD 36629 & B2V	& 0.27 & \phm{-}0.21\phn & 0.31\phn & $<$ 42 & 1 & 10583\\
HD 36879 & O8III& 0.50 & \phm{-}0.62\phn & 0.28\phn & \phn34 $\pm$ 13 & 1 & 02711s, 04654s, 04699s\\
HD 36982 & B2V	& 0.34 & \phm{-}0.050 & 0.311 & $<$ 36 & 2 & 13733, 21973\\[.05in]
HD 37022 & O6p	& 0.34 & \phm{-}0.033 & 0.186 & $<$ 16 & 2 & 07481, 13737, 13798, 14597, 15799, 19606\\
HD 37023 & B0.5V& 0.37 & -0.083 & 0.153 & $<$ 33 & 2 & 02770, 05017s\\
HD 37061 & B0.5V& 0.54 & \phm{-}0.109 & 0.044 & $<$ 27 & 2 & 07038, 08086\\
HD 37903 & B1.5V& 0.35 & \phm{-}0.384 & 0.440 & $<$ 23 & 2 & 06953, 08058, 21293, 29094\\
HD 38087 & B5V	& 0.33 & \phm{-}0.230 & 0.311 & $<$ 24 & 2 & 30046, 32748, 32749, 32750 \\[.05in]
HD 46056 & O8V 	& 0.51 & \phm{-}0.857 & 0.541 & 148 $\pm$ 37 & 2 & 08846s \\
HD 46106 & B0V	& 0.45 & \phm{-}0.591 & 0.485 & \phn55 $\pm$ 23 & 2 & 16435\\
HD 46149 & O8.5V& 0.48 & \phm{-}0.65\phn & 0.59\phn & \phn94 $\pm$ 28 & 1 & 21585, 28200\\
HD 46150 & O6V	& 0.45 & \phm{-}0.65\phn & 0.59\phn & 127 $\pm$ 19 & 1 & 18947, 18948, 21584\\
HD 46202 & O9V	& 0.47 & \phm{-}0.864 & 0.515 & 146 $\pm$ 32 & 2 & 30299 \\[.05in]
HD 46223 & O4V	& 0.54 & \phm{-}0.70\phn & 0.60\phn & 107 $\pm$ 15 & 1 & 08138s, 08338s, 08844s, 10757\\
HD 47129 & O8III& 0.36 & \phm{-}0.95\phn & 0.46\phn & \phn38 $\pm$ 10 & 1 & 10689, 13924\\
HD 48099 & O7V	& 0.27 & \phm{-}0.874 & 0.339 & $<$ 20 & 2 & 52812, 52826, 52874\\
HD 48279 & O8V	& 0.43 & \phm{-}0.72\phn & 0.47\phn & \phn60 $\pm$ 20 & 1 & 06504s\\
HD 48434 & B0III& 0.28 & \phm{-}1.10\phn & 0.52\phn & $<$ 46 & 1 & 06447s \\[.05in]
HD 73882 & O8.5V& 0.72 & \phm{-}0.788 & 0.540 & 246 $\pm$ 18 & 2 & 30076, 30077, 38299, 38311\\
HD 91824 & O7V	& 0.27 & \phm{-}0.633 & 0.473 & $<$ 87 & 2&16533 \\
HD 93028 & O9V	& 0.24 & \phm{-}0.811 & 0.166 & $<$ 69 & 2 &05521\\
HD 93204 & O5V	& 0.43 & \phm{-}0.48\phn & 0.37\phn & $<$ 32&1 &07023, 07960\\
HD 93205 & O3V	& 0.37 & \phm{-}0.45\phn & 0.50\phn & $<$ 28 & 1 & 09635s, 09655s, 09672s, 09738s\\[.05in]
HD 93222 & O7III& 0.40 & \phm{-}0.626 & 0.236 & $<$ 76 & 2&22105\\
HD 93250 & O3V	& 0.47 & \phm{-}0.58\phn & 0.50\phn & \phn70 $\pm$ 20 & 1 & 01618s, 06435s, 14746s\\
HD 93403 & O5III& 0.53 & \phm{-}0.61\phn & 0.47\phn & \phn92 $\pm$ 17 & 1 & 09075s, 09673s, 09739s\\
HD 96715 & O4V	& 0.41 & \phm{-}0.67\phn & 0.79\phn & 198 $\pm$ 19 & 1 & 22000, 43980, 43981\\
HD 147888& B3V	& 0.52 & \phm{-}0.133 & 0.339 & \phn38 $\pm$ 13 & 2 & 05159s, 05160s \\[.05in]
HD 147889& B2V	& 1.09 & \phm{-}0.151 & 0.709 & 221 $\pm$ 40 & 2 & 03924\\
HD 147933& B2V	& 0.47 & \phm{-}0.139 & 0.349 & \phn79 $\pm$ 10 & 2 & 25428, 25429, 25430, 25431, 25439\\
HD 149757& O9.5V& 0.32 & \phm{-}0.900 & 0.563 & \phn81 $\pm$ 8\phn & 2 & 36152, 36155, 36162, 36164, 36167, 38410, 41190\\
HD 152233& O6III& 0.45 & \phm{-}0.74\phn & 0.45\phn & \phn93 $\pm$ 14 & 1 & 40981, 40987, 40988\\
HD 154445& B1V	& 0.42 & \phm{-}0.309 & 0.503 & $<$ 44 & 2 & 06760s\\[.05in]
HD 162978& O8III& 0.35 & \phm{-}0.56\phn & 0.17\phn & $<$ 15 & 1 & 36945, 36948, 36949, 38302, 38303\\
HD 164816& B0V	& 0.30 & \phm{-}0.37\phn & 0.35\phn & $<$ 25 & 1 & 02814s, 15308\\
HD 165052& O7V	& 0.43 & \phm{-}0.39\phn & 0.28\phn & $<$ 18 & 1 & 15306, 17106, 45466, 45502\\
HD 167771& O7III& 0.44 & \phm{-}0.574 & 0.453 & \phn62 $\pm$ 13 & 2 & 09623s, 09633s\\
HD 168076& O4V	& 0.80 & \phm{-}0.48\phn & 0.46\phn & 127 $\pm$ 24 & 1 & 28277\\[.05in]
HD 192281& O5V	& 0.70 & \phm{-}0.77\phn & 0.43\phn & 140 $\pm$ 13 & 1 & 43147, 44636, 44637, 46539\\
HD 193322& O9V	& 0.41 & \phm{-}0.879 & 0.115 & $<$ 15 & 2 & 31326, 31327, 31328, 31329, 36935, 36936\\
HD 199579& O6V	& 0.37 & \phm{-}0.898 & 0.453 & 139 $\pm$ 13 & 2 & 31317, 31318, 31323, 31331\\
HD 200775& B2V	& 0.54 & \phm{-}0.54\phn & 0.52\phn & 276 $\pm$ 23 & 1 & 09836, 09837, 54089\\
HD 215835& O6V	& 0.65 & \phm{-}0.84\phn & 0.57\phn & \phn54 $\pm$ 34 & 1 & 45467, 45503\\[.05in]
HD 216532& O8V	& 0.86 & \phm{-}0.53\phn & 0.40\phn & 171 $\pm$ 31 & 1 & 34226\\
HD 216898& O8.5V& 0.85 & \phm{-}0.567 & 0.333 & 169 $\pm$ 34 & 2 & 43934\\
HD 217979& B2V	& 0.61 & \phm{-}0.781 & 0.580 & 297 $\pm$ 92 & 2 & 43935\\
HD 242908& O4V	& 0.60 & \phm{-}0.54\phn & 0.45\phn & $<$ 58 & 1 & 16092\\
HD 303067& B0.5V& 0.31 & \phm{-}0.733 & 0.147 & $<$ 63 & 2 & 23759\\[.05in]
HD 303308& O3V	& 0.45 & \phm{-}0.55\phn & 0.43\phn & \phn55 $\pm$ 22 & 1 & 09015s\\
BD +60 594&O9V	& 0.66 & \phm{-}0.819 & 0.411 & 103 $\pm$ 67 & 2 & 16644\\
BD +60 2522&B1V	& 0.72 & \phm{-}0.66\phn & 0.41\phn & $<$ 114 & 1 & 08840 \\
CPD $-59$ 2600&O6V& 0.53 & \phm{-}0.601 & 0.179 & $<$ 48 & 2 & 07021\\
\tablenotetext{a}{errors are 1 $\sigma$, upper limits are 2-$\sigma$}
\tablenotetext{b}{an ``s'' indicates small aperture observation.}
\tablerefs{(1) Jenniskens \& Greenberg 1993; (2)
Fitzpatrick \& Massa 1990}
\enddata
\label{table1}
\end{deluxetable}

\begin{deluxetable}{lccc}
\tablewidth{4in}
\tablecaption{ASURV results}
\tablehead{$W_\lambda$/\ebv\\versus & Spearman $\rho$ & $t$
value\tablenotemark{a} & probability\tablenotemark{b}}
\startdata
x$_0$ 	&  \phm{$-$}0.236 & \phm{$-$}1.83	& 0.072 \\
$\gamma$&  \phm{$-$}0.158 & \phm{$-$}1.21	& 0.232 \\
c$_1$	&  $-0.369$  	  & $-3.00$  	  	& 0.004 \\
c$_2$	&  \phm{$-$}0.395 & \phm{$-$}3.25	& 0.002 \\
c$_3$	&  \phm{$-$}0.203 & \phm{$-$}1.56 	& 0.123 \\
c$_4$	&  \phm{$-$}0.667 & \phm{$-$}6.76 	& $8\times10^{-9}$\\
\enddata
\tablenotetext{a}{$t = \rho \sqrt{\frac{N-2}{1-\rho^2}}$, where $N=59$
is the number of data points.}
\tablenotetext{b}{Probability that result is consistent with the null
hypothesis.}
\label{table2}
\end{deluxetable}

\newpage
\clearpage

\begin{figurehere}
\centerline{\epsfig{figure=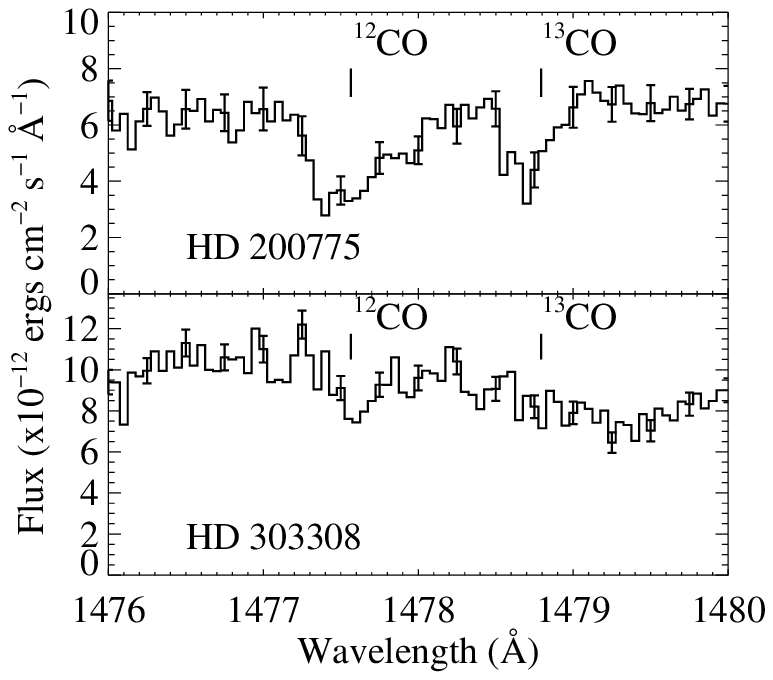}}
\caption{Sample spectra showing an example of a strong CO
$A-X$ (2--0) absorption (HD~200775) and weak (HD~303308).  The rest
wavelengths of the R(0) lines for the $^{12}$CO and $^{13}$CO are
indicated.
\label{fig1}}
\end{figurehere}

\begin{thebibliography}{}
\bibitem[\protect\citeauthoryear{{Aiello} et~al.}{{Aiello}
  et~al.}{1988}]{Aiello88}
{Aiello}, S., {Barsella}, B., {Chlewicki}, G., {Greenberg}, J.~M.,
  {Patriarchi}, P.,  \& {Perinotto}, M. 1988, \aaps, 73, 195

\bibitem[\protect\citeauthoryear{{Bloemen}}{{Bloemen}}{1989}]{Bloemen89}
{Bloemen}, H. 1989, \araa, 27, 469

\bibitem[\protect\citeauthoryear{{Boulanger}, {Prevot}, \&
{Gry}}{1994}]{Boulanger94} {Boulanger}, F., {Prevot}, M.~L., \& {Gry},
C. 1994, \aap, 284, 956

\bibitem[\protect\citeauthoryear{{Cardelli}}{{Cardelli}}{1988}]{Cardelli88}
{Cardelli}, J.~A. 1988, \apj, 335, 177

\bibitem[\protect\citeauthoryear{{Cardelli}, {Clayton}, \&
{Mathis}}{1988}] {CCM88} {Cardelli}, J.~A., {Clayton}, G.~C., \&
{Mathis}, J.~S. 1988, \apjl, 329, L33

\bibitem[\protect\citeauthoryear{{Cardelli}, {Clayton}, \&
{Mathis}}{1989}] {CCM89}
{Cardelli}, J.~A., {Clayton}, G.~C., \& {Mathis}, J.~S. 1989, \apj, 345,
245

\bibitem[\protect\citeauthoryear{{Cowie} \& {Songaila}}{{Cowie} \& 
{Songaila}}{1986}]{Cowie86} {Cowie}, L. \& 
{Songaila}, A. 1986, \araa, 24, 499 

\bibitem[\protect\citeauthoryear{{Cox} \& {Leene}}{{Cox} \&
  {Leene}}{1987}]{Cox87}
{Cox}, P.,  \& {Leene}, A. 1987, \aap, 174, 203

\bibitem[\protect\citeauthoryear{{D\'{e}sert}, {Boulanger}, \&
{Puget}}{{Desert} et~al.}{1990}]{Desert90} {D\'{e}sert}, F.~X.,
{Boulanger}, F., \& {Puget}, J.~L. 1990, \aap, 237, 215

\bibitem[\protect\citeauthoryear{{D\'{e}sert}, {Jenniskens}, \&
  {Dennefeld}}{1995}]{Desert95}
{D\'{e}sert}, F.~X., {Jenniskens}, P.,  \& {Dennefeld}, M. 1995, \aap, 303, 223

\bibitem[\protect\citeauthoryear{{Draine} \& {Lee}}{{Draine} \&
  {Lee}}{1984}]{Draine84} {Draine}, B.~T., \& {Lee}, H.~M. 1984, \apj,
285, 89

\bibitem[\protect\citeauthoryear{{Dwek} et~al.}{{Dwek} et~al.}{1997}]{Dwek97}
{Dwek}, E., et~al. 1997, \apj, 475, 565

\bibitem[\protect\citeauthoryear{{Federman} \& {Huntress}}{{Federman} \&
{Huntress}}{1989}]{Federman89} {Federman}, S.~R. \& 
{Huntress}, W.~T. , Jr. 1989, \apj, 338, 140 

\bibitem[\protect\citeauthoryear{{Federman} et~al.}{{Federman}
et~al.}{1994}]{Federman94} {Federman}, S.~R., {Strom}, C.~J., {Lambert},
D.~L., {Cardelli}, J.~A., {Smith}, V.~V., \&  {Joseph}, C.~L. 1994, 
\apj, 424, 772 

\bibitem[\protect\citeauthoryear{{Feigelson} \& {Nelson}}
{1985}]{Feigelson85} {Feigelson}, E.~D., \& {Nelson}, P.~I. 1985, \apj,
293, 192

\bibitem[\protect\citeauthoryear{{Fitzpatrick}}{{Fitzpatrick}}{1999}]{Fitzpatr%
ick99}
{Fitzpatrick}, E.~L. 1999, \pasp, 111, 63

\bibitem[\protect\citeauthoryear{{Fitzpatrick} \& {Massa}}{{Fitzpatrick} \&
  {Massa}}{1990}]{FM90}
{Fitzpatrick}, E.~L.,  \& {Massa}, D. 1990, \apjs, 72, 163

\bibitem[\protect\citeauthoryear{{Flannery}, {Roberge}, \&
  {Rybicki}}{{Flannery} et~al.}{1980}]{Flannery80}
{Flannery}, B.~P., {Roberge}, W.,  \& {Rybicki}, G.~B. 1980, \apj, 236, 598

\bibitem[\protect\citeauthoryear{{Gredel}, {van Dishoeck}, \&
{Black}}{{Gredel}, {van Dishoeck}, \&
{Black}}{1991}]{Gredel91}
{Gredel}, R., {van Dishoeck}, E. F. \& {Black}, J. H. 1991, \aap, 251, 625 

\bibitem[\protect\citeauthoryear{{Hackwell}, {Hecht}, \& {Tapia}}{{Hackwell}
  et~al.}{1991}]{Hackwell91}
{Hackwell}, J.~A., {Hecht}, J.~H.,  \& {Tapia}, M. 1991, \apj, 375, 163

\bibitem[\protect\citeauthoryear{{Hunter} et~al.}{{Hunter}
  et~al.}{1994}]{Hunter94}
{Hunter}, S.~D., {Digel}, S.~W., {De Geus}, E.~J.,  \& {Kanbach}, G. 1994,
  \apj, 436, 216

\bibitem[\protect\citeauthoryear{{Isobe} \& {Feigelson}}{1990}]{Isobe90}
{Isobe}, T., \& {Feigelson}, E.~D. 1990, \baas, 22, 917

\bibitem[\protect\citeauthoryear{{Jenniskens}}{{Jenniskens}}{1994}]{Jenniskens%
94}
{Jenniskens}, P. 1994, \aap, 284, 227

\bibitem[\protect\citeauthoryear{{Jenniskens}, {Ehrenfreund}, \&
  {D\'{e}sert}}{{Jenniskens} et~al.}{1992}]{Jenniskens92}
{Jenniskens}, P., {Ehrenfreund}, P.,  \& {D\'{e}sert}, F.~X. 1992, \aap, 265,
  L1

\bibitem[\protect\citeauthoryear{{Jenniskens} \& {Greenberg}}{{Jenniskens} \&
  {Greenberg}}{1993}]{Jenniskens93}
{Jenniskens}, P.,  \& {Greenberg}, J.~M. 1993, \aap, 274, 439

\bibitem[\protect\citeauthoryear{{Joblin}, {L\'{e}ger}, \& {Martin}}{{Joblin}
  et~al.}{1992}]{Joblin92}
{Joblin}, C., {L\'{e}ger}, A.,  \& {Martin}, P. 1992, \apjl, 393, L79

\bibitem[\protect\citeauthoryear{{Joseph} et~al.}{{Joseph}
et~al.}{1986}]{Joseph86} {Joseph}, C.~L., {Snow}, T.~P. , Jr., {Seab},
C.~G.,  \& {Crutcher}, R.~M. 1986, \apj, 309, 771 

\bibitem[\protect\citeauthoryear{{Joseph}, {Snow}, \& {Seab}}{{Joseph}
  et~al.}{1989}]{Joseph89}
{Joseph}, C.~L., {Snow}, J., T.~P.,  \& {Seab}, C.~G. 1989, \apj, 340, 314

\bibitem[\protect\citeauthoryear{{LaValley}, {Isobe}, \&
{Feigelson}}{1992}] {LaValley92}{LaValley}, M., {Isobe}, T., \&
{Feigelson}, E.~D. 1992, ASP Conf. Ser. 25: Astronomical Data Analysis
Software and Systems I, 1, 245

\bibitem[\protect\citeauthoryear{{Massa}, {Savage}, \& {Fitzpatrick}}{{Massa}
  et~al.}{1983}]{Massa83} {Massa}, D., {Savage}, B.~D., \&
{Fitzpatrick}, E.~L. 1983, \apj, 266, 662

\bibitem[\protect\citeauthoryear{{Massa} et~al.}{1998}]{Massa98}{Massa},
D., {Van Steenberg}, M. E., {Oliversen}, N., \& {Lawton}, P. 1998, in
``Ultraviolet Astrophysics Beyond the IUE Final Archive'', eds. Wamsteker and 
Riestra, ESA SP-413, p723.

\bibitem[\protect\citeauthoryear{{Mathis}, {Rumpl}, \& {Nordsieck}}{1977}]{Mathis77} {Mathis}, J.~S., {Rumpl}, W., \& {Nordsieck},
K.~H. 1977, \apj, 217, 425

\bibitem[\protect\citeauthoryear{{Mathis} \& {Cardelli}}{{Mathis} \&
  {Cardelli}}{1992}]{Mathis92} {Mathis}, J.~S., \& {Cardelli},
J.~A. 1992, \apj, 398, 610

\bibitem[\protect\citeauthoryear{{Morton} \& {Noreau}}{{Morton} \&
  {Noreau}}{1994}]{MortNor}
{Morton}, D.~C.,  \& {Noreau}, L. 1994, \apjs, 95, 301

\bibitem[\protect\citeauthoryear{{Scoville} et~al.}{{Scoville}
  et~al.}{1987}]{Scoville87} {Scoville}, N.~Z., {Yun}, M.~S., {Sanders},
D.~B., {Clemens}, D.~P., \& {Waller}, W.~H. 1987, \apjs, 63, 821

\bibitem[\protect\citeauthoryear{{Siebenmorgen} \& {Kruegel}}{{Siebenmorgen} \&
  {Kruegel}}{1992}]{Siebenmorgen92}
{Siebenmorgen}, R.,  \& {Kruegel}, E. 1992, \aap, 259, 614

\bibitem[\protect\citeauthoryear{{Tarafdar} \& {Krishna
Swamy}}{{Tarafdar} \& {Krishna Swamy}}{1982}]{Tarafdar82} 
{Tarafdar}, S.~P., \& {Krishna Swamy}, K.~S. 1982, \mnras, 200, 431 

\bibitem[\protect\citeauthoryear{{van Dishoeck} \& {Black}}{{van Dishoeck} \&
  {Black}}{1988}]{vDB88}
{van Dishoeck}, E.~F.,  \& {Black}, J.~H. 1988, \apj, 334, 771

\bibitem[\protect\citeauthoryear{{van Dishoeck} \& {Black}}{{van Dishoeck} \&
  {Black}}{1989}]{vDB89}
{van Dishoeck}, E.~F.,  \& {Black}, J.~H. 1989, \apj, 340, 273

\bibitem[\protect\citeauthoryear{{Verstraete} \& {L\'{e}ger}}{{Verstraete} \&
  {L\'{e}ger}}{1992}]{Verstraete92}
{Verstraete}, L.,  \& {L\'{e}ger}, A. 1992, \aap, 266, 513

\bibitem[\protect\citeauthoryear{{Wannier}, {Penzias}, \&
{Jenkins}}{{Wannier} et~al.}{1982}]{Wannier82} {Wannier}, P.~G.,
{Penzias}, A.~A., \& {Jenkins}, E.~B. 1982, \apj, 254, 100 


\end{thebibliography}
\end{document}